# Superconductivity in $Cu_xBi_2Se_3$ and its implications for pairing in the undoped topological insulator

Y. S. Hor[1], A. J. Williams[1], J. G. Checkelsky[2], P. Roushan[2], J. Seo[2]
Q. Xu[3], H. W. Zandbergen[3], A. Yazdani[2], N. P. Ong[2], and R. J. Cava[1]

[1]Department of Chemistry, [2]Department of Physics, Princeton University, Princeton, New Jersey 08544, USA
[3]National Centre for HREM, Department of Nanoscience, Delft Institute of Technology, 2628 CJ Delft, The Netherlands

**Abstract**

$Bi_2Se_3$ is one of a handful of known topological insulators. Here we show that copper intercalation in the van der Waals gaps between the $Bi_2Se_3$ layers, yielding an electron concentration of ~ 2 x $10^{20} cm^{-3}$, results in superconductivity at 3.8 K in $Cu_xBi_2Se_3$ for $0.12 \leq x \leq 0.15$. This demonstrates that Cooper pairing is possible in $Bi_2Se_3$ at accessible temperatures, with implications for study of the physics of topological insulators and potential devices.

Topological insulators are materials that are non-conducting in the bulk and yet display conducting electronic surface states in which a charge carrier's spin and momentum are coupled. These surface states are a distinct electronic phase of matter, with photon-like energy dispersions, and are stabilized even at high temperatures due to the topology of the system [see, e.g. 1-3]. Theoretical interest in topological surface states is very high, stimulated by their observation in HgTe-based quantum wells[4,5], but especially since the prediction[6,7] and then observation[8] that they are present on the surface of bulk Bi-Sb alloy crystals. Topological surface states have recently been observed in a second bulk materials class, $Bi_2Se_3$ and $Bi_2Te_3$[9,10]. Several schemes have been proposed to search for novel electronic excitations of the surface states, particularly Majorana fermions[11], which are potentially useful for topological quantum computing [e.g. 12-18]. All the proposed schemes rely on the opening of an energy gap in the surface state spectrum by inducing superconductivity through the proximity effect. However, Cooper pairing of electrons at accessible temperatures in a topological insulator has never been reported. Here we show that the topological insulator $Bi_2Se_3$ can be made into a superconductor by Cu intercalation between the $Bi_2Se_3$ layers. This implies that Cooper pairing can occur in $Bi_2Se_3$ up to about 4 K. Due to their intrinsic chemical and structural compatibility, electronic junctions between $Bi_2Se_3$ and $Cu_xBi_2Se_3$ are feasible. Such junctions are very promising for investigating novel concepts in physics as well as for electronic devices based on topological surface states.

$Bi_2Se_3$ is a classical layered compound[19], made from double layers of $BiSe_6$ octahedra (Figure 1(a)). The resulting Se-Bi-Se-Bi-Se five layer sandwich is only weakly van der Waals bonded to the next sandwich, through the outer Se layers, yielding a material with excellent basal plane cleavage and excellent quality layered crystals both in the bulk and in thin films[19-21]. This layered nature results in the fact that both substitutional and intercalative chemical manipulations are possible. The dopant employed here, Cu, may either intercalate between the Se layers, as it does in $Cu_xTiSe_2$[22], or randomly substitute for Bi within the host structure, as has been reported for the NaCl structure compound $CuBiSe_2$[23]. This dual nature was recognized early on in Cu doping studies of $Bi_2Se_3$[24,25], where substantial differences in the electrical properties of Cu-substituted $Bi_{2-x}Cu_xSe_3$ and copper intercalated $Cu_xBi_2Se_3$ were reported[26] and it was concluded that Cu acts as an ambipolar dopant. Whether Cu is incorporated by intercalation or substitution in $Bi_2Se_3$ is nominally controlled by the elemental ratios, but because both kinds of incorporation are possible, the precise distribution of Cu between intercalated and substituted sites is expected to be strongly processing dependent, a factor that we find to be a significant influence on the presence of the superconductivity.

Single crystals of $Cu_xBi_2Se_3$ were grown by melting stoichiometric mixtures of high purity elements Bi (99.999 %), Cu (99.99 %) and Se (99.999 %) at 850 ºC overnight in sealed evacuated quartz tubes. The crystal growth took place via slow cooling from 850 to 620 ºC and then quenching in cold water. The resultant crystals are easily cleaved along the basal plane leaving a silvery shining mirror like surface. The silvery surfaces turn golden after one day exposure to air, suggesting that exposure to air should be kept to a minimum. $Bi_{2-y}Cu_ySe_3$ crystals were grown by the same technique.

The crystal structure of the Cu intercalated $Cu_xBi_2Se_3$ crystals was confirmed by X-ray powder diffraction using a Bruker D8 diffractometer with Cu Kα radiation and a graphite diffracted beam monochromator. Transmission electron microscopy (TEM) analysis was performed with a Philips CM300UT electron microscope operating at 300 kV. Specimens for the TEM studies were obtained by crushing the crystals under ethanol to form a suspension and then by dripping a droplet of this suspension onto a holey carbon film on a Cu grid, and by making cross sections of single crystals using ion milling. Scanning tunneling microscopy measurements were performed in a home-built cryogenic scanning tunneling microscope (STM) operated at 4.2 K in ultrahigh vacuum. The samples were cleaved *in situ* to expose a pristine surface.

Resistivity and AC magnetization measurements were performed in a Quantum Design physical property measurement system (PPMS). The standard four-probe technique, employing silver paste contacts cured at room temperature, was used for resistivity measurements, with the electric currents applied in the basal plane of the crystals. Hall Effect measurements were in a home-built apparatus 0.3 and 15 K. A Quantum Design Superconducting Quantum Interference Device (SQUID) was used to measure DC magnetization in a 10 Oe applied magnetic field, and AC susceptibility data were taken in the PPMS with a magnetic field of amplitude 5 Oe.

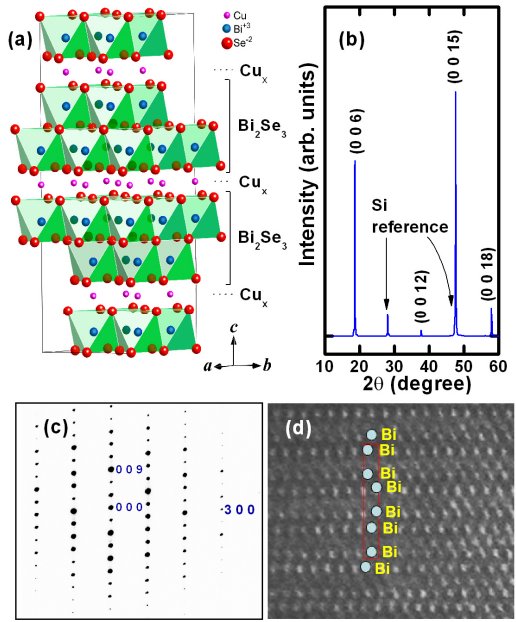

**Figure 1. Structural characterization of superconducting $Cu_{0.12}Bi_2Se_3$.** (a) The crystal structure of Cu intercalated $Bi_2Se_3$. (b) X-ray diffraction scan showing the *00L* reflections from the basal plane of a cleaved $Cu_{0.12}Bi_2Se_3$ single crystal with Si powder diffraction as a calibration. (c) The <110> zone axis electron diffraction pattern for $Cu_{0.12}Bi_2Se_3$. (d) High resolution electron microscope image of a representative area of $Cu_{0.12}Bi_2Se_3$, showing the regular array of layers (labeled by atom type) and the absence of stacking defects on the nanoscale.

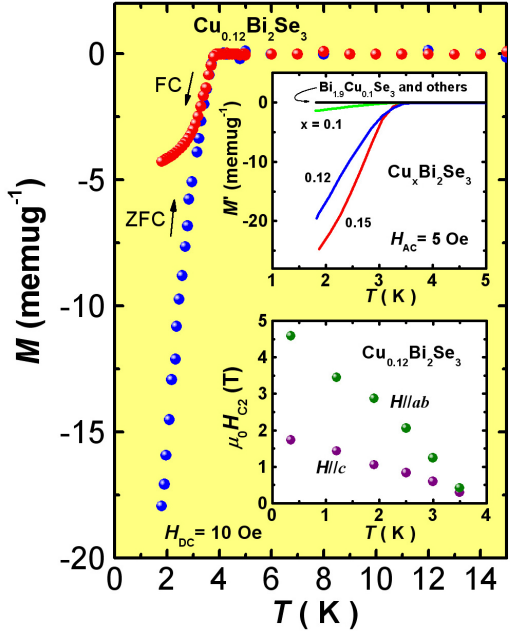

**Figure 2**. **Magnetization characterization of the superconducting transition.** The temperature dependent magnetization of a single crystal of $Cu_{0.12}Bi_2Se_3$ shows a superconducting transition $T_c \sim 3.8$ K in zero field cooled (ZFC) and field cooled (FC) measurements. Upper inset: superconductivity occurs only in a narrow window of x in $Cu_xBi_2Se_3$. Superconductivity is not found for x < 0.1 and x > 0.3, or in $Bi_{2-y}Cu_ySe_3$, $Cu_2Se$, $CuBi_3Se_5$, $BiCuSe_2$, and $BiCu_3Se_3$ (data labeled in inset as "$Bi_{1.9}Cu_{0.1}Se_3$ and others"). Lower inset: temperature dependence of the superconducting upper critical field, $H_{c2}$(T), for a single crystal of $Cu_{0.12}Bi_2Se_3$ for the magnetic field applied parallel to the c-axis and parallel to the ab-plane.

Single crystals of $Bi_{2-x}Cu_xSe_3$ prepared for x = 0 to 0.15 were never superconducting. In contrast, single crystals of $Cu_xBi_2Se_3$ within the composition range 0.10 < x < 0.15 were reproducibly superconducting. The crystallographic quality of the superconducting single crystals was investigated on both the long-range scale, by X-ray diffraction, and the nanoscale, by electron diffraction and high resolution lattice imaging. An X-ray diffraction (XRD) scan of the basal plane reflections from a single crystal of superconducting $Cu_{0.12}Bi_2Se_3$ is shown in Figure 1(b). This scan shows several of the *00L* reflections obeying the expected systematic absences for the rhombohedral space group of $Bi_2Se_3$[19]. The reflections are sharp, indicating good crystalline quality on the long range. For all $Cu_xBi_2Se_3$ crystals, we observe a subtle increase in the c-axis lattice parameter (the layer stacking direction), for example from $c \sim 28.67$ Å for undoped $Bi_2Se_3$ to $c \sim 28.73$ Å for $Cu_{0.12}Bi_2Se_3$. We infer, by analogy to similar behavior in observed related layered selenides[22], that the intercalated Cu in the van der Waals gap partially occupies the octahedrally coordinated 3b (0, 0, ½) sites in the $R\bar{3}m$ space group, shown in Figure 1(a). Electron diffraction (Figure 1(c)) and High Resolution Electron Microscopy (Figure 1(d)) in zones including the c* stacking axis, taken from many fragments of $Cu_xBi_2Se_3$, showed no sign of stacking faults or intergrowth, indicating that the quality of the crystals is good on the nanoscale. None of the diffraction experiments (or the STM experiments, described below) indicated the presence of long range or short range ordering of the Cu in the interstitial sites.

Magnetic characterization of the superconducting transition in one of the single crystals of $Cu_{0.12}Bi_2Se_3$ is shown in the main panel of Figure 2. The superconducting transition temperature is approximately 3.8 K. The magnetization observed at the lowest temperature of the zero field cooled (ZFC) measurement on the single crystal is about 20 % of that expected for full diamagnetism, but that estimate represents a lower limit because temperatures substantially below the transition were not accessible in our apparatus and the transition is not complete at the temperature where the field is applied for the ZFC measurement. Superconductivity was observed only in a narrow window of x in $Cu_xBi_2Se_3$, for $0.1 \leq x \leq 0.3$, with the optimal single crystal compositions between x = 0.12 and x = 0.15 (inset Figure 2). Small deviations of Se stoichiometry from 3.00/formula unit suppress the superconductivity. For nominal x greater than 0.15 in $Cu_xBi_2Se_3$ melts, pure single crystals were not obtained – the melts included minor amounts of impurity phases. These minor extra phases were $Cu_2Se$ and $CuBi_3Se_5$ (also reported as $Cu_{1.6}Bi_{4.8}Se_8$[26]). Samples of these phases, and other reported phases in the Cu-Bi-Se system, $CuBiSe_2$[23] and

BiCu$_3$Se$_3$[27] as well as any compositions deviating from the stoichiometry Cu$_x$Bi$_2$Se$_3$ for $0.1 \leq x \leq 0.30$ were not superconducting above 1.8 K (upper inset Figure 2). The temperature dependence of the upper critical field, $H_{c2}$, determined from resistivity (see below), is shown in the lower inset of Figure 2. At 0.3 K, $H_{c2}$ is 1.7 T and 4.6 T for the field parallel to the *c*-axis, $H//c$, and parallel to the *ab* ("basal") plane, $H//ab$, respectively. Cu$_x$Bi$_2$Se$_3$ is a strongly type II superconductor with a Ginzburg-Landau parameter $\kappa \sim 50$. The inferred coherence lengths ($\xi_c = 52$ Å, $\xi_{ab} = 140$ Å) imply a pair condensate with a moderately large anisotropy ($\sim 3$).

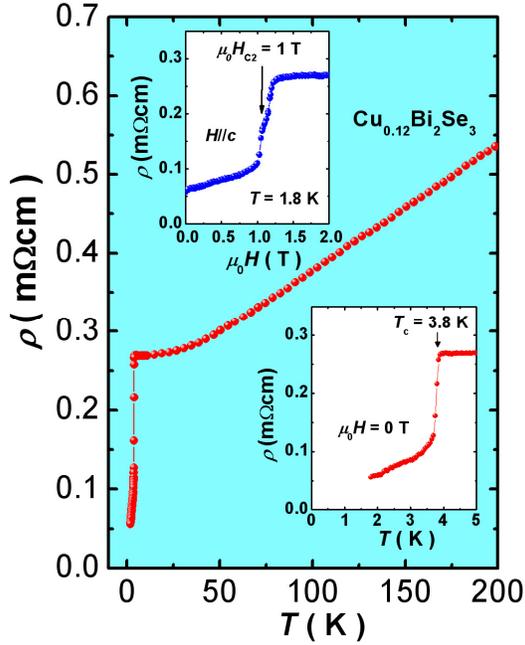

**Figure 3. The temperature dependent resistivity of a Cu$_{0.12}$Bi$_2$Se$_3$ single crystal.** The resistivity behavior of a Cu$_{0.12}$Bi$_2$Se$_3$ crystal with applied current in the *ab*-plane. The lower inset shows that the superconducting transition occurs at ~ 3.8 K. The upper inset shows the magnetoresistance plot at $T = 1.8$ K for the field applied parallel to the *c*-axis. An upper critical field of $\mu_oH_C \sim 1$ T is observed.

Figure 3 shows the temperature dependence of the resistivity for a single crystal of Cu$_{0.12}$Bi$_2$Se$_3$, measured in the *ab* plane. The resistivity is weakly metallic, dropping from approximately 0.5 mΩcm to 0.27 mΩcm on cooling to 4 K. Hall effect measurements indicated that this crystal is *n*-type, with a temperature independent (between 4 and 300 K) carrier density of approximately $\sim 2 \times 10^{20}$ cm$^{-3}$, comparable to what has been observed previously for Cu$_x$Bi$_2$Se$_3$[25]. This carrier concentration is one order of magnitude higher than is found in native Bi$_2$Se$_3$[24,25], and two orders of magnitude higher than is found for crystals with chemical potentials tuned by Ca doping[28,29]. The superconducting transition, which is expanded in the lower inset of Figure 3, occurs at ~ 3.8 K. The resistivity does not drop to zero below $T_c$ however, indicating that there is not a continuous superconducting path in the crystal. This behavior was seen in many crystals and is contrary to what is generally seen in superconductors, where a continuous zero resistance path can often be seen even when the superconducting volume fraction is relatively small. We attribute this to the sensitivity of the superconducting phase to processing and stoichiometry, as discussed further below. Resistivity data showing the suppression of superconductivity in an applied field (upper inset Figure 3) was employed to determine the $H_{c2}(T)$ behavior presented in Figure 2.

How the Cu intercalant gives rise to superconductivity in Bi$_2$Se$_3$ is of particular interest. To investigate this, we studied the (001) surface of a cleaved Cu$_{0.15}$Bi$_2$Se$_3$ single crystal using an STM at 4.2 K. Several distinct features can be identified in the STM topographic images. Figure 4 shows topographies of filled and unoccupied states over an area of size 250 Å by 250 Å. One feature (type 1), with an apparent height of ~ 2 to 3 Å, occurs in different shapes and sizes, and appears as red or orange regions in Figure 4. From the height of these features and their shapes, we conclude that they are located on the cleaved surface and are clusters of intercalated Cu atoms. A second feature (type 2) appears as a bright triangular area, yellow in Figure 4, of lateral dimension ~ 20 Å. We associate these features with intercalated Cu in sub-surface van der Waals gap layers. They have a symmetry-defined shape. Hence there is no evidence of Cu cluster formation within the van der Waals layers below the exposed surface, suggesting that the clusters of Cu atoms on the surface have formed by surface diffusion. Neither of these features is present in native or Ca-doped $Bi_2Se_3$ [28] and thus they are clearly associated with the Cu intercalation. Because these topographic features have identical appearance in both filled and unoccupied state topographies in the STM images, we conclude that Cu is not uniquely a donor or acceptor in this system. A third feature (type 3), also having a three-fold symmetry, (visible for example as a triangle of dark circles in the +1 V image in Figure 4), but with bias dependence, is also present. Unambiguous identification of this feature will require further study.

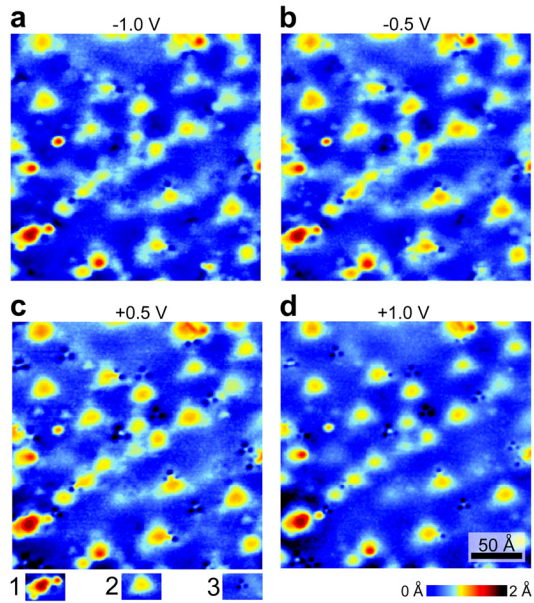

**Figure 4. STM topography of the $Cu_{0.15}Bi_2Se_3$ (001) surface.** STM topographic images over a 250 Å by 250 Å area showing the filled states at bias voltages of (a) -1.0 V and (b) -0.5 V; and the unoccupied states at bias voltages of (c) +0.5 V and (d) +1.0 V. Three types of topological features are identified, denoted as 1, 2, and 3 in the legend. Type 1 and type 2 features are identified as Cu on the cleaved surface and intercalated in van der Waals layers beneath the surface, respectively.

Previous experimental and theoretical work on $Bi_2Se_3$ suggests that it may be the most interesting of the bulk topological insulators known to date because it can in fact be made resistive in the bulk through doping and also because only topological surface states are present[28-31]. We have here shown that $Cu_xBi_2Se_3$ displays superconductivity at accessible temperatures in the doping range $0.12 < x < 0.15$, adding an important attribute to the favorable characteristics of this system.

Most proposals to manipulate the topological surface states in $Bi_2Se_3$ rely on the opening of a superconducting gap by proximity to a superconductor. Our result, showing that pairing exists up to 3.8 K (albeit at a density of $2 \times 10^{20}$ cm$^{-3}$), provides strong impetus to these efforts. As $Cu_xBi_2Se_3$ and $Bi_2Se_3$ are chemically closely similar, supercurrents should readily tunnel between them. Furthermore, if modulation doping is achieved by atomic-scale control of the Cu sites (e.g. by molecular-beam epitaxy), we anticipate novel device structures based on having superconducting layers alternating with topological insulators. In addition to superlattice structures, we envision "inverted" structures, in which two topological insulators sandwich between them a single superconducting $Cu_xBi_2Se_3$ layer. In a magnetic field, the two surfaces are connected by Abrikosov vortices acting as "worm holes". Such schemes introduce notions that should expand research on the topological surface states.

## Acknowledgements

The work at Princeton was supported by the National Science Foundation MRSEC program, grant DMR-0819860.